\documentclass[12pt]{article}

\usepackage[utf8]{inputenc} 
\usepackage[T1]{fontenc}    
\usepackage{hyperref}       
\usepackage{url}            
\usepackage[style=numeric, backend=biber]{biblatex}
\usepackage{authblk}
\usepackage[margin=2.75cm]{geometry}

\begin{document}

\title{Noumenal Labs White Paper: How To Build A Brain}
\author[1,2]{Maxwell J. D. Ramstead}
\author[1]{Candice Pattisapu}
\author[1]{Jason Fox}
\author[1,3]{Jeff Beck}
\affil[1]{Noumenal Labs, USA}
\affil[2]{Queen Square Institute of Neurology, University College London, UK}
\affil[3]{Department of Neurobiology, Duke University, USA}

\date{\today}

\maketitle

\begin{abstract}
This white paper describes some of the design principles for artificial or machine intelligence that guide efforts at Noumenal Labs. These principles are drawn from both nature and from the means by which we come to represent and understand it. The end goal of research and development in this field should be to design machine intelligences that augment our understanding of the world and enhance our ability to act in it, without replacing us. In the first two sections, we examine the core motivation for our approach: resolving the grounding problem. We argue that the solution to the grounding problem rests in the design of models grounded in the world that we inhabit — not mere word models. A machine super intelligence that is capable of significantly enhancing our understanding of the human world must represent the world as we do and be capable of generating new knowledge, building on what we already know. In other words, it must be properly grounded and explicitly designed for rational, empirical inquiry, modeled after the scientific method. A primary implication of this design principle is that agents must be capable of engaging autonomously in causal physics discovery. We discuss the pragmatic implications of this approach, and in particular, the use cases in realistic 3D world modeling and multimodal, multidimensional time series analysis.
\end{abstract}

\subsection*{Acknowledgments}
We thank Karl Friston for useful feedback. 

\newpage

\section{Introduction: How to build a brain}

In this white paper, we describe the design principles for artificial or machine intelligence that guide efforts at Noumenal Labs, drawn from nature and the means by which we come to represent and understand it. We believe that the end goal of research and development in this field should be to design machine intelligences — models, agents, and networks — that augment our understanding of the world and enhance our ability to act in it, without replacing or displacing us. We argue that such systems must be able to understand the world in the same way that we do — so we can understand each other. To this end, we are pioneering a new approach to the design of machine intelligence, to build machines that think like us. 

The first principle is that the design of machine intelligence must emulate the laws of self-organization that govern the behavior of intelligent agents in the natural world, and which led to the emergence of human intelligence, over evolutionary and developmental time. Here, we draw on Bayesian mechanics and the free energy principle \cite{ramstead2023bayesian, friston2023simpler}, which tell us that anything that continues to exist in a given environment must come to represent or model that environment. In other words, any agent that successfully persists in an ecological niche comes to form a model of that niche — otherwise it would not exist \cite{friston2011embodied}. This line of reasoning implies that the structure of our representation of the world must mirror the structure of our world. For simple unicellular life, this means that a cell must effectively model the molecular soup in which it lives. For larger, more sophisticated organisms (such as ourselves), this means that "successful" agents — those that persist over time — effectively model the macroscopic world in which they live, a world composed of a wide variety of other macroscopic objects, including other agents \cite{veissiere2020thinking, lake2017building}. A core implication of this is that the intelligence that we observe in the natural world is not a result of a generic statistical learning paradigm or a one size fits all algorithm.

Rather, intelligent behavior in nature is a highly specialized product of the situation or ecological niche in which that intelligence lives. The primary implication of this natural intelligence or embodied intelligence argument is that a human-like intelligence should be grounded in a model of the world in which humans live \cite{pezzulo2024generating}. A common grounding in a shared representational format makes it possible to compare and share representations of the world, and thus meaningfully understand and communicate with each other \cite{collins2024building}. This suggests that the underlying models that are employed by machine intelligence should be world models based in the world in which we live, i.e. models that represent the structure of the physical world, the objects that populate it, and the dynamic relations among them.

The second principle is that the design of machine intelligence must conform to the structure of scientific explanation, i.e., core models should be grounded in human intuitive physics, but enhanced by scientific precision. This is where the two sources of inspiration — nature and the means by which we come to understand it — intersect. While a basic correspondence between the structure of human intuitive understanding of the world is necessary for the design of machines that think like us, the special properties of scientific investigation affords a rational, empirical extension of this intuition. To achieve this kind of machine intelligence requires endowing agents with models that are based upon our best causal, object-centered, relational models of the macroscopic world in which we live. Accordingly, we propose that machine intelligence that thinks like us should be built, not from 'neurons', but by composing simple object centered physical models that form the atomic elements of our natural thought processes.  

The structure of this white paper is as follows. In the first two sections, we examine the core motivation for the approach: the grounding problem. We point out that state of the art machine learning architectures are essentially word models, and that the failure modes of these models suggest that a fundamentally new approach to machine intelligence is required. We then build on the principal argument from Bayesian mechanics — that any agent necessarily must come to model the structure of their environment — to argue that the models that ground human understanding of the world are epitomized by a specific class of macroscopic physical models. This implies that a prerequisite for sophisticated human-like intelligence is a set of highly structured, composable models, based on macroscopic physics rather than ‘neurons’ or rote statistical learning. We also argue that requiring artificial agents to use models that are consistent with our intuitive understanding of the world enables them to ‘think like us’ and to produce reductionist (simplified, object centered) explanations for their behavior that are both accurate and understandable via analogical and methodologically reductionist reasoning.

In the following section, we argue that a machine super intelligence, one that is capable of enhancing our understanding of the human world by generating new knowledge from what we already know, should be explicitly endowed with rational, empirical models of the sort employed in scientific inquiry. We argue that scientific explanations are a formalization of our intuitive approach to gaining an understanding of the world and, as a result, they lead to simplified explanations that improve our understanding. This class of models and explanations also enable the kind of systems engineering thinking and analogical reasoning that we apply when conceptualizing increasingly complex situations. We use the remainder of this section to elucidate the fundamental properties of scientific explanations that we believe should motivate our design choices. Generically, models should enable data prediction and data compression, and afford simplified, object-centered explanations of the data. That is, models do not merely predict data, they follow a reductionist or object-centered explanatory modeling strategy, explaining complex phenomena by breaking them up into their component parts and their causal interactions at multiple hierarchically nested scales or levels — much as we would do in a quantitative discipline such as physics. Finally, and most crucially, they should be composable, in the sense that it should be possible to sensibly combine any number of object-centered models into a many-object scene model, at any scale.

Finally, we review the implications of our approach for the design of artificial agents that have the ability to act in the world. The key insight here pertains to the difference between a model and an agent, and how that difference relates scientific explanation and empirical inquiry. While a model passively processes data, an agent interacts with the structure that generates that data. We believe that the ideal artificial agents should not just use scientific models, but also be designed to behave like scientists. Namely, they should be capable of autonomously engaging in rational and empirical reasoning by efficiently generating and testing hypotheses and following the principles of optimal experimental design, i.e., active learning and active inference. Crucially, such machine intelligence of this sort not only continuously learns, but also, continuously generates new models, both \textit{de novo} and by composing existing models into new, more expressive ones. This capability allows agents to maximally exploit a causal, object-centered modeling approach and ultimately enables the kind of systems engineering thinking that gave rise to the age of modern technological achievement. 

\section{The grounding problem and word models}

A core motivation for our design philosophy is known as the grounding problem \cite{harnad1990symbol}. If we want to design and build super-intelligent machines that think like us and understand us and each other in the same way that we do, then we need to build machines that think in the same terms as we do — machines whose representations are anchored in the same foundation as ours, or grounded in the same domain as ours. In other words, the common ground between human and artificial agents, that we seek to understand, and by which we want to be understood, must be a common representational format. This is a prerequisite for authentic alignment between human and machine intelligences.

Today’s artificial intelligence systems are best described as \textit{word} models, rather than \textit{world} models \cite{wong2023word}. State of the art machine learning models are grounded in a space of linguistic representations. In other words, they map or embed large multimodal and multidimensional datasets onto a space of labels — that is, onto words. But, despite appearances, and in spite of arguments to the contrary \cite{fodor1975language}, human language is both a poor representation of the physical world and a poor representation of human thought. Language is not necessary to represent and interact with the world, indeed the vast majority of species on the planet do just fine without language. Moreover, language is not a reliable representation of our thought processes. As anyone who has performed a cognitive or psychological experiment knows, linguistic self-reported explanations for behavior are one of the least reliable forms of data, and often amounts to little more than post-hoc rationalization \cite{mercier2017enigma}. More problematically, language is at least three degrees removed from the reality upon which our models are actually grounded. As a result, it affords, at best, imprecise descriptions of reality. Scientific progress in understanding the structure of the world was achieved only by creating a mathematical language that afforded more precise descriptions of reality than did our internal models of reality. 

What are the implications for the design of machine intelligence? While human natural language very clearly has its uses — human language has been a major driver of human evolution, as well as social and cultural progress — it is a poor foundation upon which to build a machine that represents the world like we do \cite{pezzulo2024generating}. All this calls into question the dominant approach of modern artificial intelligence based on next word prediction. And so we propose an alternative approach that synthesizes intuitive and mathematical understanding — harnessing them for the design of machine intelligence, in a way that augments human intelligence.

\section{From word models to grounded world models}

If language does not provide a proper grounding for (machine) intelligence, then what does? We believe that machine intelligence should be grounded in a world model, based on our understanding of the physical world \cite{pezzulo2024generating, lake2017building}. This approach to the design of machine intelligence is premised on the idea that natural intelligences, like those displayed by living creatures with sophisticated brains, must have evolved an accurate understanding of the world in which they live — indeed, that their success, their very existence, requires it. 

This is not mere speculation — but the result of an information theoretic analysis. This argument is premised on the primary result from recent work on Bayesian mechanics and the free energy principle \cite{ramstead2023bayesian, friston2023simpler}. This analysis demonstrates that any self organizing system that persists through time must come to form a representation or model of the environment in which it exists. In the special case of sophisticated living organisms, the implication is that any organism that successfully persists in a given ecological niche must have come to form a representation or model of that niche \cite{friston2011embodied}. For example, a unicellular creature must have a model of the relevant aspects of the chemistry of its environment to inform the way in which it should respond to any changes in that environment. Otherwise, it would simply cease to exist. Similarly, more macroscopic organisms, like human beings, must have generated an understanding of the macroscopic objects that exist in their environment and whose behavior is governed, for example, by Newton’s laws of motion. 

The implication of this is that the atomic elements of our thought processes are inherited from the properties of the simple physical systems that govern the interactions between the objects the world in which we evolved. Thus, an intelligence that thinks like we do must be grounded in an object-centered, intuitive physics like our own \cite{lake2017building}. 

By “object-centered”, we mean that a sensible representation of the world is one that carves it up into more or less discrete objects, which are defined both by their intrinsic properties and how they causally interact with one another \cite{battaglia2013simulation}: A rock that is thrown travels in their air following a ballistic trajectory, until it hits the water, which creates a ripple. More sophisticated systems are simply the result of a large number of these causal interactions being composed or chained together, as in a Rube Goldberg machine. To be successful, brains had to evolve an understanding of this interactional and compositional structure of the world \cite{battaglia2016interaction}. Indeed, arguably, the key to the vastly expanded capabilities of human beings is a result of the ability to represent objects and object types compositionally, thereby allowing us to consider the consequences of novel combinations of objects and properties. 

Indeed, as brains evolved, they vastly augmented their ability to represent and act in the world by repurposing and reusing basic object-centered models of physical processes in increasingly complex ways \cite{anderson2010neural, anderson2021after}. The structural evolution of the mammalian cortex followed the same trajectory. The remarkable capabilities of the human brain were similarly achieved by incrementally expanding our capabilities by dynamically repurposing and recombining the functionality of simpler cortical regions, endowing it with the expressiveness required to adapt to more complex systems \cite{hawkins2021thousand}. 

The magic was in the compositional nature of our models and cortical structures, but the starting point was an understanding of the physical world composed of objects that interact via Newton’s laws \cite{lake2017building}. This explains the remarkable quality of our intuition regarding physical systems, even from a very young age \cite{spelke1990principles, spelke1995development}. It also elucidates why a common means by which we explain complex phenomena is via analogy to simpler physical systems, e.g., trajectories over space-time curvature are like a heavy ball rolling on a rubber mat \cite{lakoff2008metaphors}. 

It is worth noting that, in the light of Bayesian mechanics, it is unsurprising that this compositional feature of cortical evolution emerged, because it mirrors the compositional structure of complex phenomena in the macroscopic world in which we live. Similarly, it is unsurprising that our understanding of the world has been formalized via scientific reductionism. More precisely, this is a methodological reductionism \cite{ayala1974studies, ayala1987biological}, one that reduces complex phenomena into causally interacting components, each of which is described by a simpler physical model. This is not an ontological reductionism, which would reduce all emergent processes to the more ‘basic’ physical processes from which they emerge. It is also unsurprising that our greatest technological achievements resulted from inverting this process, and building increasingly useful machines by composing them from well understood parts. Indeed, the compositionality of both the world and our thought processes is ultimately what enables creative thinking, systems engineering, novel invention, and arguably, life itself. 

The above considerations suggest that machine intelligence that thinks like us should be built not from ‘neurons’, but rather from atomic elements of thought — which, we argue, consist of the simple physical models that describe the world in which we learned to survive. Our more sophisticated models are then built by combining these atomic elements of thought, in service of solving novel problems via a process similar to program synthesis. This too is not a mere assertion or speculation, since the principal driver of the evolution of emergent high-level intelligence is the need to adapt dynamically to a world of perpetually increasing complexity, which calls for increasingly expressive models \cite{anderson2021after}. This approach constitutes a significant departure from standard neural networks, for which the fundamental computational unit is a ‘neuron’ — moving to one in which the fundamental computational unit is a sophisticated dynamical system or model of a physical system upon which our experience is based \cite{hawkins2021thousand, lake2017building}.

\section{What is the structure of a grounded world model?}

So far, we have argued that, if we want to design a machine intelligence that thinks in the same way that we do, then we should be focused less on function approximation and more on building composable models of the world in which we live. Of course, we only experience the world indirectly and our sensory experience can be unreliable, but we have access to a very good explanatory model of the world in which we live, specifically, the one provided by scientific investigation and rational empirical inquiry. 

Here, we can invert the Bayesian mechanics argument to conclude that our best scientific models of macroscopic phenomena underlie our understanding of our world. Bayesian mechanics tell us that our representation of the world must have evolved to mirror its structure. As a result, we can use the structure of our most successful models of the world — the structure of scientific models of macroscopic phenomena and systems engineering — as a proxy for the structure of understanding of it and thus the ideal structure of the understanding of artificial agents. Thus, it is our view that an ideal foundation for machine intelligence — the type of representation or embedding that it uses — must share the structure of our best scientific models of the natural world. This is not merely because our scientific models of the natural world happen to be the best models of the natural world that we have. It is also because our best scientific models are formalizations of our natural thought processes — and thereby, our best scientific models and our natural thought processes must share the same ground; namely, the physical world. As the famous English biologist and anthropologist Huxley put it, "Science is simply common sense at its best, that is, rigidly accurate in observation, and merciless to fallacy in logic" \cite{huxley1880crayfish}. Indeed, scientific thinking and rational empiricism are not modern inventions, but are better understood as a formalization or extension of core thought processes that we have inherited from our interactions with a highly structured world. 

So, what is the structure of our scientific understanding of the macroscopic word and how does it lead to design principles for machine intelligence that enhance human understanding? 

Generically, a good scientific model has the following key properties:
\begin{itemize}
    \item Prediction and explanatory data compression
\end{itemize}

A good scientific model of phenomena in our macroscopic world has these additional properties: 
\begin{itemize}
    \item Simplification afforded by an explicit object-centered and relational representation
    \item Dynamic, causal, and sparse representations of interactions
    \item Compositional or composable, meaning relations are expressed in a common format
\end{itemize}

\subsection{Prediction and explanatory data compression}

In scientific investigation, the word “model” usually denotes a parameterized statistical model, or a mathematical structure that is used to represent relationships (ideally causal relationships) between variables, some of which are observed outcomes (i.e., data) and others unobserved, that is, hidden or latent. Pragmatically, models serve two functions: prediction of future data and explanatory data compression. 

Prediction is what makes models useful, by enabling planning, counterfactual reasoning, and decision making (e.g., in applications such as autonomous drones and robotics). But data compression is what makes models insightful and enables them to contribute to our understanding of the world. Data compression, if done properly, is explanatory in that it provides a simplified representation of the underlying data. Simplification is key here: A good explanation is one that compactly summarizes the relationships in the data using a small number of variables while maintaining predictive accuracy. 

This difference between mere prediction (or function fitting) and explanation is key here. It is very different for a model to be able to predict the next data point or the label associated with an image, and for a model to provide a summarized and simplified explanation regarding why the outcome turned out the way that it did that is consistent with our understanding of the world. State of the art generative AI has focused almost exclusively on prediction, with genuinely impressive results. Generative AI algorithms, for example, are designed to discover and encode sophisticated correlations between linguistic terms (labels) in enormous datasets and exploit them for next token prediction or image prediction. This is accomplished by utilizing massively over parameterized models needed to make gradient descent learning work. But the emphasis on rote prediction and reliance on over parameterized models means that successful prediction has come at the expense of the kind of explainability that reduces the complexity of the data into understandable chunks. 

Scientific explanations, on the other hand, are explicitly reductionist in the methodological sense that they seek to explain phenomena by dividing them up into component parts, each of which can be described with a very simple model \cite{ayala1974studies}. This is dimensionality reduction applied to data. It is the process of explaining complex phenomena as being the result of the interaction of simple, well understood phenomena. This means that a methodologically reductionist explanation (1) carves the world up into understandable ‘chunks’, or objects of different types that behave in simple, easy to understand ways, (2) they compactly represent the effects of interactions between objects using a common format that allows for the prediction of the outcomes of novel interactions. In technical parlance, we say that reductionist models of this sort are object-centered and relational. The principal advantage of this approach is that it enables one to not only decompose a world into a set of interacting objects, it also enables the kind of systems engineering that enables us to predict the properties of novel combinations of objects and exploit that knowledge to build something completely new from an explicit understanding of the functioning of the parts.

\subsection{Simplification afforded by an explicit object-centered representation}

It is important to note that methodological reductionism taken to an extreme is not what we are talking about. We are not proposing modeling a basketball as a collection of polymer chains that are made airtight by weakly interacting molecular bonds. The argument from Bayesian mechanics applies at the scale of our perception and affordances. We conceptualize the world via a kind of object-centered reductionism that generates semantic interpretations based upon behavior and interactions at the macroscopic level. So a basketball is a thing that bounces in a certain way and can be held or thrown. Thus, the simple description of the world that we seek is one that compresses the observation of each surface element of the ball to a simple equation that governs its dynamics. A successful description of this kind has the property that knowledge of the granular details has been rendered irrelevant when predicting the behavior of the object. In the literature on emergent phenomena, this is known as downward causation and is a critical element a good explanation \cite{barnett2023dynamical}. For example, a volume of an ideal gas is composed of a very very large number of particles, but its effects on its surroundings can be summarized by just three variables: temperature, pressure, and density. It makes sense to label a volume of gas as an object precisely because labeling a collection of particles as a volume of gas allows us to better predict its behavior and interactions with other things. That is, given these three quantities, knowledge of the precise positions of each of the particles is irrelevant to the prediction of the effects of the gas volume on its environment. 

This allows us to treat a volume of gas as a simple object. We can generalize and formalize this notion in mathematical terms by defining an object or object type in terms of the union of inputs and -- known in the statistics literature as a Markov blanket \cite{pearl1998}. The Markov blanket is simply the statistical signature of an object: it is the set of variables that statistically separate an object from its environment \cite{friston2013life, karl2012fepforbiological, sakthivadivel2022weak}. Crucially, it is the statistics of the Markov blanket — its behavior at the boundary, where it interacts with its environment — that determine the object type. As a result, a model of the boundary of an object is equivalent to a model of the object as a whole, regardless of the fine details of that model and the goings on inside the object. This leads naturally to a blanket-based taxonomy, wherein object type is effectively determined entirely by its interactions with its environment. Since that environment is made of other objects, those interactions are relational, in the sense that they pertain to specific object types.

\subsection{Simplification afforded by a sparse, relational representation of interactions}

We just said that Markov blankets afford a simplified representation of objects defined by their relations to other objects. This motivates an object-centered representation of the world, but is only part of the story. A good object-centered segmentation of the world does not merely simplify our representation of objects; it also simplifies our representation of the interactions between objects. This is accomplished by choosing an object-centered segmentation of the world that reduces the number of interactions which must be considered at any given time and represents those interactions via simple low dimensional channels. 

In more formal terms, the structure of the causal interactions of objects with each other is sparse, dynamic, and relational. It is sparse because objects only interact infrequently or with a small number of other objects. It is dynamic because the set of objects with which a given object interacts can change over time. And it is relational because the kinds of interactions between objects are specific to the type of objects involved in the interaction. This dynamic sparsity greatly simplifies our models by compressing the set of possible interactions into understandable chunks between objects that are also understandable chunks.

\subsection{Simplified generalization via compositionality}

The final element of this approach to dynamical systems segregation is compositionality. A naive object-centered and relational model has to have an understanding of how every pair of objects interacts. In a world composed of many objects this requires modeling every kind of possible interaction. Taken to an extreme, this results in a situation wherein it is impossible to predict how two objects would interact if they have never been seen together. 

In physics and systems engineering this issue is resolved by requiring that object interactions be expressed in a common language. For mechanical systems that language is forces, for fluids or electrical systems that language its fluxes. This constitutes a third class of restrictions on the models that we may consider and allows one to take two well understood objects and instantly generalize how they would interact with one another, despite having never observed them interacting. In Bayesian mechanics, force exertion is generalized, and treated as an information transfer of a particular kind. We represent and model the interaction of simple objects with one another in the format of classical mechanical forces. As objects become more complex, the “forces” that they exert on each other accordingly become more complex. This is because the information transfers that they exert on each other become more complex. The Markov blanket based approach allows us to characterize these ‘forces’ in a simplified manner, in terms of the effects that things have at their boundary. 

\subsection{Automation of causal physics discovery}

Formalizing these notions and implementing them into world models requires embedding a great deal of structure into the model that we are using. The Markov blanket construction supplies the structure of the atomic elements of the world model, while the world model itself is composed of a large set of blanket based elements and the rules that govern their interactions. In this setting, a given scene consists of some unknown number of objects, each of which has an abstract embedding that models its current state (i.e., a position, velocity, shape, color, texture) and its type. Coupled to this state space representation is a dynamic graph that specifies which objects are interacting at a given point in time, i.e., a sparse adjacency matrix. This explicit object-centered approach generalizes previous approaches to physics discovery that have been successfully applied in complex fluid dynamical settings. The critical difference is that in most physics discovery situations, the objects are assumed to be observed and are all of the same type and related in the same way. The Markov blanket construction allows us to generalize and abstract this approach to a consideration of multiple different kinds of objects, each of which explains a subset of the observables. 

This transforms the problem of macroscopic physics discovery — the problem of discovering macroscopic objects and their behavior from data — into a problem of finding a partition of the system into subsystems, following the line of reasoning described above. Namely, a partition is good if it affords a simplified description of the underlying dynamics, in terms of labels associated with object membership. Moreover, it can be extended to include hierarchical structure, effectively partitioning a microscopic system into a set of objects which are themselves sets of objects. The focus on boundary dynamics also reduces the compute requirements necessary to model sophisticated systems. Ultimately, the object identification and typification that we have described allows us to automate the process of discovering an object centered physics directly from data, by imposing the structure of our understanding of the natural world onto our models.  

\section{From causal scientific intervention to agent design}

Causal physical models are pragmatically useful because they instruct us about where and how we can intervene within a system to learn something about its behavior or drive it to a desired endpoint — that is, they provide a means for control over physical systems. Object-centered models augment this by directing our attention toward interventions that affect the system at the scale at which we live, making the actions of agents (both human and artificial ones) more impactful.

The point of building world models is that they are the prerequisite ingredient needed to design agents that effectively act in a world. In the context of artificial intelligence, the difference between an agent and a model is that a model represents data (in the sense that it explains, predicts, or compresses data), whereas an agent, in addition, interacts actively and adaptively with the process that generates data. Learning about the world and its structure is an active process that requires interaction with the world. Humans don’t just perceive in order to act, they act in order to perceive \cite{merleau2013phenomenology, varela2017embodied}, as in the visual saccades that our eyes deploy to palpate the world visually. Scientific investigation is an extension of this sensorimotor loop: Scientists don’t just model data, they use intervention and design experiments to both exploit and discover causal structure, taking actions that enhance their understanding of how the world works . 

An idealized machine intelligence should be capable of expanding our understanding of the natural world and, therefore, should be explicitly designed for rational inquiry via causal interventions that enhance explanations for data as well as decisions. Without such a causal model, scientific experimentation would be a blind process of trial and error. With causal models to guide intervention, agents can explicitly engage in actions that improves the quality of their world models. This is the critical value add of active inference approaches over model based reinforcement learning. It is also how a scientist behaves: A scientist takes actions that have been explicitly designed to resolve uncertainty of their beliefs about the causal structure of the world. More specifically, scientists seek out interventions that are likely to generate new data that tests their causal models.

Taken together, this provides an absolutely crucial element for the design of machine (super) intelligence. We want models and agents that explicitly encode the objects and object types that are relevant to a domain, along with their intrinsic properties, and their relations and interactions with other objects and object types. We want machine intelligences that learn the rules that govern the behavior of these objects directly from data in an unsupervised manner. We want machine intelligence that can engage in causal physics discovery, and that are designed to reason and act like little scientists, generating actionable causal insights. We want automated interactive discovery. We want systems that actively discover the structure of their world and that are able to take appropriate actions in order to resolve uncertainty in their beliefs. 

\section{What we get from grounded world models}

We have described the design principles that we are pursuing and the core features that make a good world model — and the ensuing approach to causal physics discovery, allowing us to identify high level objects, object types, and object relationships directly from data in an unsupervised fashion. This raises the question: in practice, what do we get from causal object-centered world models? While the areas of application for this technology are numerous, we focus on a few key areas to give a sense of the opportunity that is afforded by this technology. In general, any area where the extraction of discrete objects and strategies matters will be an exciting testbed for this technology. 

Generically, the approach to world modeling via physics discovery that we have described is ideal for modeling high dimensional, multimodal time series data. An immediate and exciting area of application is rapid modeling of realistic environments — for use cases such as video game asset generation and virtual robotics training. When applied to physical systems or 3D world data, the kind of physics discovery technology that we are developing outputs objects that are easily recognizable as video game assets. Indeed, extracting objects based on their blanket statistics is a way of extracting entire, structured objects and their properties for a game world. This is not merely a bounding box and associated textures — but rather, an identified object, classified into an object type according to its interaction profile with other objects using a generalized notion of ‘interaction’ that is a simple abstraction of the underlying forces. That is, the key upshot is that assets generated in this way come complete with simplified dynamics, which generalize the idea of force vectors that are used in the gaming industry. Indeed, video game assets operate on force vectors and simulate classical mechanics under the hood, which makes the simulation of complex macroscopic objects. This explicit use of force vectors often results in the kind of stiff dynamics that can cause a ball wedged between a foot and the floor to launch away at unrealistic velocities. Because our macroscopic physics discovery approach generalizes the notion of force based upon observation, it naturally smoothes out these kinds of idiosyncrasies and enables a data driven real-world-to-video-game asset mapping that relies on simplified discovered dynamics. The upshot is an automatic video game asset discovery engine that models phenomena in a frugal and efficient way.

Another obvious area of application is financial analysis and investment. Financial data is made up of huge amounts of raw stock prices sampled every few seconds, fundamental corporate data, and analysts reports. This is an ideal test case, because the relevant data forms a time series that operates on multiple timescales, but is ultimately generated by complex networks of causal factors. Now, in principle, one could fit all that data in a brute force fashion to a neural network — and it might even generate useful predictions into the behavior of the market. But as we described above, this approach would not provide us with actionable insight or elucidate the possibility of sophisticated plays based on complex interactions between assets and opinions. Identifying causal relationships between groups of time series simplifies the problem of predicting future outcomes greatly, and gives us the ability to identify specific causal relationships driving the data, which can be used to guide the search for the most predictive factors for the behavior of a given asset or class of assets. 

\section{Conclusion: How to build a grounded world model}

The aim of this paper was to introduce the design principles for artificial or machine intelligence that we are pioneering at Noumenal Labs. We argued that the north star of research and development in artificial intelligence should be the design of machine intelligences that understand the world in the same way that we do — so we can understand each other. More specifically, Noumenal Labs is designing models and agents equipped with grounded world models — that is, models that explicitly encode the objects and object types that pertain to a specific domain of application, and the rules that govern their causal interactions. Such machine intelligences are designed to discover and maximally leverage a causal, object-centered world model. This common ground between machine and human intelligences enables mutual understanding and alignment — and ultimately, enables and automates the kind of systems engineering thinking that gave rise to the age of modern technological achievement.

\printbibliography

\end{document}